\documentclass[a4paper,twoside,11pt]{article}

\usepackage{BeaCed_article}

\title{Bipartite dimer representation of squared 2d-Ising correlations}
\author{B\'eatrice de Tili\`ere
\thanks{{\small 
UPMC Univ Paris 06, UMR 7599,
Laboratoire de Probabilit\'es et Mod\`eles Al\'eatoires, 4 place Jussieu, 
F-75005 Paris.}
{\small\texttt{beatrice.de\_tiliere@upmc.fr.}}
{\small Supported by the ANR Grant 2010-BLAN-0123-02.}
}}
\date{}

\begin{document}

\maketitle

\begin{abstract}
The Bozonisation identities of \cite{Dubedat} show that squared 2d-Ising order and disorder correlations are equal to 
$\pm$ the ratio of bipartite dimer partition functions. In this self-contained paper, 
we give another proof of these identities using the approach of \cite{BoutillierdeTiliere:XORloops}. Our proof is 
more direct and allows to keep track of order and disorder in XOR-Ising configurations.


\end{abstract}

\section{Introduction}\label{sec:1}

Let $\Gs=(\Vs,\Es)$ be a finite, planar embedded graph. Consider the Ising model on the graph $\Gs$ with coupling constants 
$\Js=(\Js_e)_{e\in\Es}$, and denote by $\Zising(\Gs,\Js)$ the Ising partition function. 

\begin{figure}[ht]
  \begin{center}
\psfrag{e}[c][c]{\scriptsize $e$}
\psfrag{1}[c][c]{\scriptsize $1$}
\psfrag{G}[c][c]{$\Gs$}
\psfrag{GQ}[c][c]{$\GQ$}
\psfrag{u1}[c][c]{\scriptsize $u_1$}
\psfrag{u2}[c][c]{\scriptsize $u_2$} 
\psfrag{u3}[c][c]{\scriptsize $u_3$} 
\psfrag{u4}[c][c]{\scriptsize $u_4$} 
\psfrag{f1}[c][c]{\scriptsize $f_1$}
\psfrag{f2}[c][c]{\scriptsize $f_2$}
\psfrag{f3}[c][c]{\scriptsize $f_3$}
\psfrag{f4}[c][c]{\scriptsize $f_4$}
\psfrag{f5}[c][c]{\scriptsize $f_5$}
\psfrag{f6}[c][c]{\scriptsize $f_6$}
\psfrag{Je1}[c][c]{\scriptsize $\bar{\Js}_e=\Js_e+i\frac{\pi}{2}$}
\psfrag{Je2}[c][c]{\scriptsize $\bar{\Js}_e=-\Js_e$}
\psfrag{Je3}[c][c]{\scriptsize $\bar{\Js}_e=\Js_e$}
\psfrag{w1}[c][c]{\scriptsize $\cosh^{-1}(2\Js_e)$}
\psfrag{w4}[c][c]{\scriptsize $\tanh(2\Js_e)$}  
\psfrag{w2}[c][c]{\scriptsize -$\cosh^{-1}(2\Js_e)$}
\psfrag{w5}[c][c]{\scriptsize $\tanh(2\Js_e)$}  
\psfrag{w3}[c][c]{\scriptsize $\cosh^{-1}(2\Js_e)$}
\psfrag{w6}[c][c]{\scriptsize -$\tanh(2\Js_e)$} 
\psfrag{weights}[c][c]{\scriptsize Weights $\nu(\bar{\Js})$} 
\includegraphics[width=\linewidth]{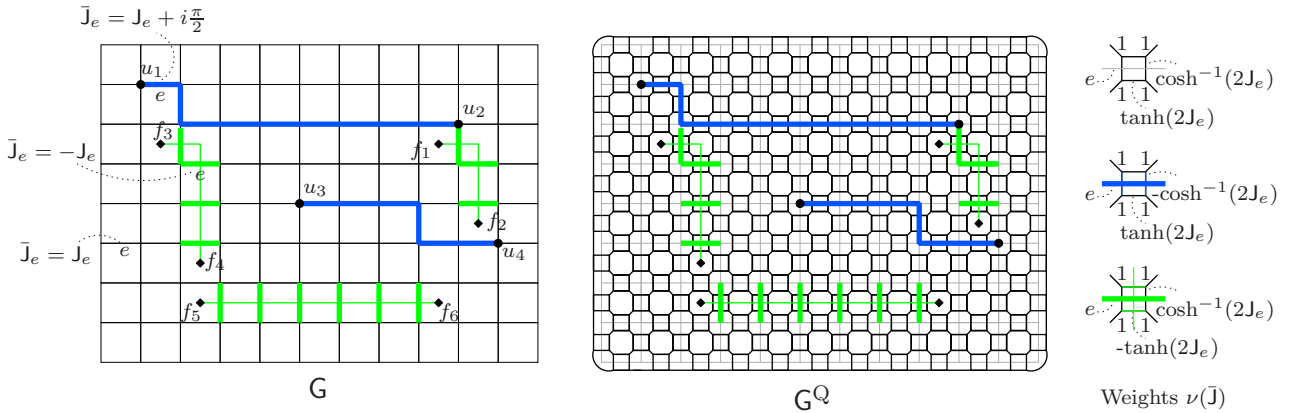}
    \caption{Left: order and disorder in the Ising model on $\Gs$. Center:
    Image of order and disorder in the bipartite graph $\GQ$. Right: dimer weight function $\nu(\bar{\Js})$
    associated to the modified coupling constants $\bar{\Js}$, expressed as a function of the original coupling constants 
    $\Js$.}\label{fig:FigIsingDefect2}
  \end{center}
\end{figure}

Following Kadanoff and Ceva
\cite{KadanoffCeva}, we introduce \emph{order} and \emph{disorder} in the model: \emph{order} amounts to adding $i\frac{\pi}{2}$ to coupling constants 
along $n$ paths of the graph $\Gs$ joining $2n$ vertices $u_1,\dots,u_{2n}$, see Figure \ref{fig:FigIsingDefect2} (left, blue paths); 
\emph{disorder} amounts to negating coupling constants of dual edges of $m$ paths of the dual graph $\Gs^*$ joining $2m$ faces $f_1,\dots,f_{2m}$ of $\Gs$,
see Figure \ref{fig:FigIsingDefect2} (left, green paths). Denote by $\bar{\Js}=(\bar{\Js_e})_{e\in\Es}$ the modified coupling constants, 
and by $\mathord{<}\sigma_{u_1}\dots\sigma_{u_{2n}}\ \mu_{f_1}\dots\mu_{f_{2m}}\mathord{>}_{(\Gs,\Js)}$ the Ising correlation defined as the ratio 
$\Zising(\Gs,\bar{\Js})/\Zising(\Gs,\Js)$.

Consider the dimer model on the finite, planar, bipartite graph $\GQ=(\VQ,\GQ)$ constructed from $\Gs$, see Figure 
\ref{fig:FigIsingDefect2} (center). Suppose that edges of $\GQ$ are assigned the weight function 
$\nu(\Js)=(\nu(\Js))_{\es\in\EQ}$, defined in Figure 
\ref{fig:FigIsingDefect2} (top right), and denote by $\Zdimer(\GQ,\nu(\Js))$ the corresponding dimer partition function.
Consider also the modified weight function $\nu(\bar{\Js})$ obtained from the 
modified coupling constants $\bar{\Js}$: $\nu(\bar{\Js})$ is defined as in Figure \ref{fig:FigIsingDefect2} (top right,
with $\Js$ replaced by $\bar{\Js}$); expressing $\nu(\bar{\Js})$ as a function of the coupling constants $\Js$ yields Figure 
\ref{fig:FigIsingDefect2} (right: top, middle and bottom). Written with the notations of this paper, the Bozonisation
identities of \cite{Dubedat} are stated as follows:
\begin{thm}[\cite{Dubedat}]\label{thm:main}
The squared Ising correlation $\mathord{<}\sigma_{u_1}\dots\sigma_{u_{2n}}\ \mu_{f_1}\dots\mu_{f_{2m}}\mathord{>}_{(\Gs,\Js)}^2$ is equal to $\pm$ the following ratio of bipartite dimer 
partition functions:
\begin{equation}\label{equ:main}
 \mathord{<}\sigma_{u_1}\dots\sigma_{u_{2n}}\ \mu_{f_1}\dots\mu_{f_{2m}}\mathord{>}_{(\Gs,\Js)}^2=
 (-1)^{|\Gamma|}\frac{\Zdimer(\GQ,\nu(\bar{\Js}))}{\Zdimer(\GQ,\nu(\Js))},
\end{equation}
where $|\Gamma|$ is the number of edges in paths defining order (the blue paths of Figure \ref{fig:FigIsingDefect2}).
\end{thm}

\begin{rem}$\,$\vspace{-0.2cm}
\begin{itemize}
 \item In proving Theorem \ref{thm:main} we use the approach of \cite{BoutillierdeTiliere:XORloops},
which has the advantage of starting from two independent Ising models living on the \emph{same} graph, instead of
 one on the \emph{primal} and one on the \emph{dual} graph as in \cite{Dubedat}. 
 An outline of the two methods is given in Remark 1.4. of \cite{BoutillierdeTiliere:XORloops}. 
 We believe that our approach is more transparent since it does not require 
 the use of Kramers and Wannier's duality \cite{KramersWannier1, KramersWannier2} to
 transform Ising correlations of the dual graph into those of the primal graph.
 Moreover, it provides a coupling with the \emph{XOR-Ising model}, also known as the \emph{polarization} of the model, 
 obtained by taking the product of the spins of two independent Ising models, see \cite{IkhlefRajabpour,Picco,WilsonXOR}. 
 In proving Theorem \ref{thm:main}, we keep track of the effect of order and disorder in the XOR-Ising model.
 
 \item Theorem \ref{thm:main} is stated for the Ising model with free boundary conditions. In Section~\ref{sec:22} we explain how,
 by transforming the graph and keeping it planar, and possibly adding disorder, all boundary conditions enter the framework of free
 boundary ones.
 \item Consequences of Theorem \ref{thm:main} are expressions as ratio of dimer partition functions for: squared, Ising 
 spinor variables correlations, spin correlations, and magnetization. This is explained in Section \ref{sec:4}.
 \item By \cite{KadanoffCeva} order and disorder correlations satisfy Kramers and Wannier's duality:
 \begin{equation*}
(-1)^{|\Gamma|}\mathord{<}\sigma_{u_1}\dots\sigma_{u_{2n}}\ \mu_{f_1}\dots\mu_{f_{2m}}\mathord{>}_{(\Gs,\Js)}=
(-1)^{|\Gamma^*|}\mathord{<}\sigma_{f_1}\dots\sigma_{f_{2m}}\ \mu_{u_1}\dots\mu_{u_{2n}}\mathord{>}_{({\Gs}^*,{\Js}^*)},
 \end{equation*}
where edges of the dual graph ${\Gs}^*$ are assigned \emph{dual coupling constants} ${\Js}^*={\Js}^*(\Js)$, defined by:
${\Js}^*=\bigl({\Js}^*_{e^*}=-\frac{1}{2}\ln(\tanh\Js_e)\bigr)_{e^*\in\Es^*}.$
In particular, when the graph $\Gs$ has no disorder, order correlations of $\Gs$ are mapped to disorder correlations of the
dual graph $\Gs^*$, with dual coupling constants. Kramers and Wannier's duality can also be seen in the numerator and denominator of the
right-hand-side of \eqref{equ:main}. This is a consequence of the following two facts:
modified coupling constants also
satisfy the duality relation, \emph{i.e.}, $\overline{{\Js}^*}_{e^*}=-\frac{1}{2}\ln(\tanh\overline{\Js}_e)$; and
$\cosh^{-1}(2{\Js}^*_{e^*})=\tanh(2 \Js_e)$, for all choices of coupling constants~$\Js$.
\item The advantage of having an expression involving the dimer partition function of the bipartite graph $\GQ$, is that it is equal to the determinant
  of the \emph{Kasteleyn matrix} of~$\GQ$, which is a weighted, oriented adjacency matrix of the graph
  \cite{Kasteleyn1,Kasteleyn2,TF}.
\end{itemize}
\end{rem}

\medskip

\begin{center}
\textsc{Outline}
\end{center}
\begin{itemize}
 \item \underline{Section \ref{sec:2}}. Definition of the Ising model, of order and disorder. Treatment of other boundary conditions. 
 Definition of the dimer model on the bipartite graph $\GQ$.
 \item \underline{Section \ref{sec:3}}. Proof of Theorem \ref{thm:main}. Observing that modified Ising weights $\bar{\Js}$ can be seen as another
 choice of coupling constants, Theorem \ref{thm:main} is a consequence of the results of \cite{BoutillierdeTiliere:XORloops},
 the proof is thus rather short. We nevertheless outline the different steps since it allows us to keep track of order and 
 disorder in XOR-Ising
 configurations, and makes this paper self contained. Note that disorder (not order) was already considered 
in \cite{BoutillierdeTiliere:XORloops}; but we did not interpret it as a possible choice of coupling constants,
rather we wrote it as a function of the original coupling constants and were interested in its effect on the homology of polygon configurations
arising from high and low temperature expansions \cite{KramersWannier1,KramersWannier2}.
Also, disorder is absent from the final statements of \cite{BoutillierdeTiliere:XORloops}, it is only present in the intermediate steps involved in handling surfaces of genus $g$.
\item \underline{Section \ref{sec:4}}. Consequences of Theorem \ref{thm:main} for squared, Ising spinor variables correlations, spin correlations
and magnetization.
\end{itemize}

\emph{Acknowledgements.} We would like to thank Cédric Boutillier, Dmitry Chelkak, David Cimasoni and Adrien Kassel for their interest
in expressing squared Ising spin correlations using the approach of \cite{BoutillierdeTiliere:XORloops}.

\section{Definitions and boundary conditions}\label{sec:2}

\subsection{Two-dimensional Ising model, order and disorder}\label{sec:21}

Consider a finite, planar graph $\Gs=(\Vs,\Es)$, together with a collection of positive \emph{coupling constants} 
$\Js=(\Js_e)_{e\in \Es}$ indexed by edges of $\Gs$. 
The \emph{Ising model on $\Gs$, with coupling constants~$\Js$}, is defined as follows.
A \emph{spin configuration} $\sigma$ is a function of the vertices of $\Gs$ taking values in $\{-1,1\}$. The probability 
on the set of spin configurations $\{-1,1\}^\Vs$, is 
given by the \emph{Ising Boltzmann
measure} $\PPising$, defined by:
$$
\forall\, \sigma\in\{-1,1\}^\Vs,\quad\PPising(\sigma)=\frac{1}{\Zising(\Gs,\Js)}\exp\Bigl(\sum_{e=uv\in
\Es}\Js_e\sigma_u\sigma_v\Bigr),
$$
where $\Zising(\Gs,\Js)=\sum_{\sigma\in\{-1,1\}^\Vs}
\exp\bigl(\sum_{e=uv\in \Es}\Js_e\sigma_u\sigma_v\bigr)$
is the normalizing constant, known as the \emph{Ising partition function}. 

It is convenient to consider the graph $\Gs$ as embedded in the sphere.
Suppose that the embedding of the dual graph $\Gs^*$ is such that dual vertices are in the interior of the faces of $\Gs$, and 
such that primal and dual edges cross exactly once. Following Kadanoff and Ceva \cite{KadanoffCeva}, we introduce \emph{order} and 
\emph{disorder} in the system.
Given positive integers $n$ and $m$, 
let $u_1,\dots,u_{2n}$ be $2n$ vertices of $\Gs$ and
$f_1,\dots,f_{2m}$ be $2m$ vertices of the dual graph $\Gs^*$. Consider $n$ loop-free paths $\gamma_1,\dots,\gamma_n$ of $\Gs$, 
such that $\gamma_j$ has endpoints $u_{2j-1}$, $u_{2j}$, and $m$ loop-free paths 
$\gamma_1^*,\dots,\gamma_m^*$ of $\Gs^*$, such that 
$\gamma_j^*$ has endpoints $f_{2j-1}$, $f_{2j}$, see Figure \ref{fig:FigIsingDefect2} (left).
Denote by $\Gamma$ the set of edges of the paths
$\gamma_1,\dots,\gamma_n$, and by $\Gamma^*$ the set of edges dual to edges of the paths 
$\gamma_1^*,\dots,\gamma_m^*$. Note that $\Gamma^*$ is a subset of edges of the primal graph $\Gs$.
Define the following modified coupling constants $\bar{\Js}=(\bar{\Js}_e)_{e\in\Es}$:
\begin{equation*}\label{equ:modifiedIsingweights}
\forall\ e\in\Es,\quad\bar{\Js}_e=
\begin{cases}
\Js_e+i\frac{\pi}{2}&\text{ if }e\in\Gamma\\
-\Js_e&\text{ if } e\in\Gamma^*\\ 
\Js_e&\text{ otherwise}.
\end{cases}
\end{equation*}
Then, $\Zising(\Gs,\bar{\Js})=\sum_{\sigma\in\{-1,1\}^\Vs}
\exp\bigl(\sum_{e=uv\in \Es}\bar{\Js}_e\sigma_u\sigma_v\bigr)$ is the
corresponding \emph{modified Ising partition function}.

\begin{rem}
If only order or only disorder is introduced in the system, the modified Ising partition function is independent
of the paths $\Gamma$ or $\Gamma^*$. If both order and disorder are considered, then changing the paths might induce a sign change 
\cite{KadanoffCeva}. It will be convenient to suppose that the 
embedded planar graph $\Gs$ and its dual $\Gs^*$ are such that the paths in $\Gamma$ and $\Gamma^*$ can be chosen to be pairwise 
disjoint. This is the only assumption we make on the graphs $\Gs$ and $\Gs^*$; examples use a piece of $\ZZ^2$ simply because it is
easier to draw.
\end{rem}

\subsection{Boundary conditions}\label{sec:22}

The Ising model introduced in Section \ref{sec:21} is also known as the Ising model with \emph{free-boundary conditions}. We now
discuss how to handle other boundary conditions. Since the graph $\Gs$ is embedded in the sphere, fixing
boundary conditions amounts to fixing spins on boundary vertices of a face $\Fs$ of $\Gs$. 
We suppose that boundary edges of the face $\Fs$ are not covered by edges of $\Gamma$ or $\Gamma^*$.

\begin{figure}[ht]
  \begin{center}
\psfrag{G}[c][c]{$\Gs$}
\psfrag{Gp}[c][c]{$\Gs'$}
\psfrag{F}[c][c]{$\Fs$}
\psfrag{merging}[c][c]{\scriptsize merging}
\includegraphics[width=12cm]{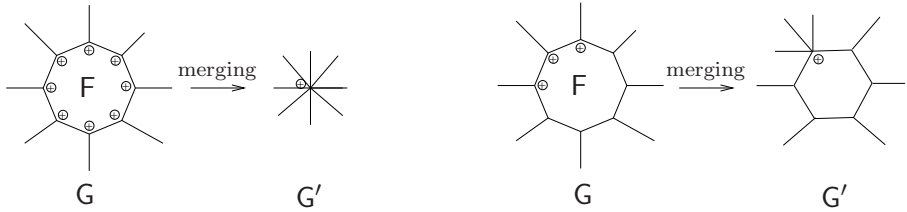}
    \caption{Merging of vertices and edges to handle plus-boundary conditions (left) and plus-free-boundary conditions (right).}
    \label{fig:FigBoundary2}
  \end{center}
\end{figure}

Consider first \emph{plus-boundary conditions}, meaning that all spins on boundary vertices of $\Fs$ are +1. Denote 
by $\Es_{\partial \Fs}$ the set of boundary edges of the face $\Fs$. Then,
up to the constant\footnote{
the constant does not depend on the modified coupling constants $\bar{\Js}$ because, by assumption, boundary edges of $\Fs$ are not covered by $\Gamma$
and $\Gamma^*$} $\prod_{e\in\Es_{\partial \Fs}}e^{\Js_e}$, the modified Ising partition function is equal to the one of the graph $\Gs'$ obtained from $\Gs$ by merging the face
$\Fs$ into a single vertex, and where this vertex is fixed to having spin $+1$, see Figure \ref{fig:FigBoundary2} (left).
Since the modified partition function is invariant under the transformation $\sigma \leftrightarrow -\sigma$, it is up to a factor
$\frac{1}{2}$, the modified partition function of the graph $\Gs'$ with free boundary conditions. The graph $\Gs'$ is also planar and
embedded in the sphere, so that it enters the framework of this paper. A mixture of plus and free-boundary conditions can be handled in a similar way, by contracting all edges with fixed +1 spins,
see Figure \ref{fig:FigBoundary2} (right).

\begin{figure}[ht]
  \begin{center}
\psfrag{-Je}[c][c]{\scriptsize -$\Js_e$}
\psfrag{G}[c][c]{$\Gs$}
\psfrag{Gp}[c][c]{$\Gs'$}
\psfrag{Gpp}[c][c]{$\Gs^{''}$}
\psfrag{F}[c][c]{$\Fs$}
\psfrag{merging}[c][c]{\scriptsize merging}
\includegraphics[width=12cm]{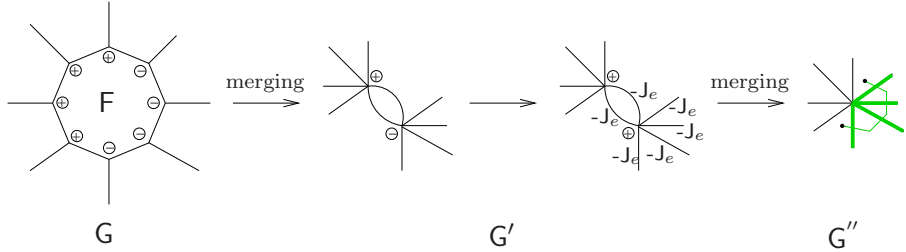}
    \caption{Merging of vertices and edges and introduction of a disorder line to handle Dobrushin boundary conditions.}
    \label{fig:FigBoundary1}
  \end{center}
\end{figure}

Consider now \emph{Dobrushin boundary conditions}, meaning that the boundary of the face $\Fs$ is split into two connected
components, one having +1 spins and the other -1 spins. Up to a constant, the modified Ising partition function is equal to the one of
the graph $\Gs'$ obtained by merging all vertices and edges of the face $\Fs$ having respectively
+1 spins and -1 spins, see Figure \ref{fig:FigBoundary1} (center). Let $u$ be the vertex with fixed -1 spin, then the 
modified partition function of $\Gs'$ is equal to the one where the spin at $u$ is +1 and 
coupling constants on edges incident to $u$ are negated. We now have an Ising model with two vertices on the boundary
of a face of degree 2 with fixed +1 spins. Up to a constant, the modified Ising partition function is equal to the one of the graph 
$\Gs^{''}$ obtained by merging the two vertices into a single vertex with +1 spin, and adding a disorder line, 
see Figure \ref{fig:FigBoundary1} (right). Up to a constant $\frac{1}{2}$ it is equal to the modified
partition function of the graph $\Gs^{''}$ with free boundary conditions,  
and enters again te framework of this paper.

\subsection{Dimer model on the bipartite graph $\GQ$}\label{sec:23}

\begin{figure}[ht]
  \begin{center}
\includegraphics[width=14.5cm]{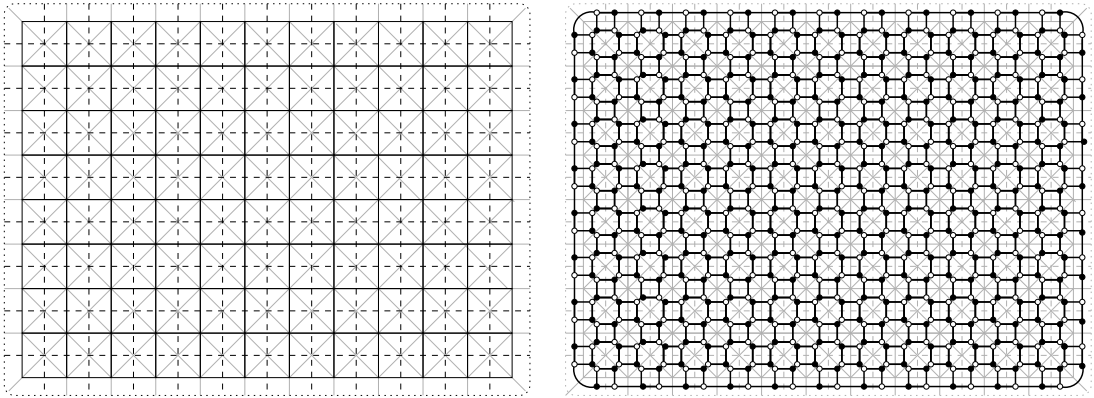}
    \caption{Left: planar embedding of the graph $\Gs$ (plain black lines), dual graph $\Gs^*$ (wide-dotted black lines), the 
    tiny-dotted line is a spreadoutway of representing the vertex of $\Gs^*$ corresponding to the outer-face (in the planar embedding) of $\Gs$, 
    and the quad-graph $\GQ$ (grey lines). Right: the bipartite graph $\GQ$ (plain black lines).}
    \label{fig:FigBipartite}
  \end{center}
\end{figure}

The bipartite graph $\GQ=(\VQ,\EQ)$ is constructed from the graph $\Gs$ and its dual $\Gs^*$ as follows. Let us first define
the \emph{quad-graph}, denoted $\Gquad$, whose vertices
are those of $\Gs$ and of the dual graph $\Gs^*$. A dual vertex is
then joined to all primal vertices on the boundary of the corresponding
face. The embedding of $\Gquad$ is chosen such that its edges do not intersect
those of $\Gs$ and $\Gs^*$, see Figure
\ref{fig:FigBipartite} (left, grey lines).
Consider the graph obtained by superimposing the primal graph $\Gs$, the dual graph $\Gs^*$,
the quad-graph $\Gquad$, and by adding a vertex at the crossing of each primal
and dual edge. Then, the dual of this graph, denoted by $\GQ$, is the graph on which the dimer model lives, 
see Figure \ref{fig:FigBipartite} (right). It is bipartite
and consists of \emph{quadrangles} and \emph{legs} connecting the quadrangles, legs are crossing edges of the quad-graph $\Gquad$. 
In each quadrangle, two edges are ``parallel'' to an edge $e$ of $\Gs$ and two edges are ``parallel'' to its 
dual edge $e^*$ of~$\Gs^*$.

Suppose that edges of $\GQ$ are assigned a positive weight function $\nu=(\nu_\es)_{\es\in\GQ}$.
The \emph{dimer model on $\GQ$ with weight function $\nu$}, is defined as follows.
A \emph{dimer configuration} of $\GQ$, also known as a \emph{perfect matching}, is a subset of edges $\Ms$ of $\GQ$ such that
every vertex is incident to exactly one edge of $\Ms$, see Figure \ref{fig:FigIsingDefect5}. 
Let us denote by
$\M(\GQ)$ the set of dimer configurations of the graph $\GQ$. The probability on the set of dimer configurations $\M(\Gs)$ is 
the \emph{dimer Boltzmann measure} $\PPdimer$, defined by:
\begin{equation*}
\forall\,\Ms\in\M(\Gs),\quad\PPdimer(\Ms)= \frac{\prod_{\es\in M}\nu_{\es}}{\Zdimer(\GQ,\nu)},
\end{equation*}
where $\Zdimer(\GQ,\nu)=\sum_{\Ms\in\M(\GQ)} \prod_{\es\in M}\nu_{\es}$ is the normalizing constant known as the \emph{dimer partition function}.

\section{Proof of Theorem \ref{thm:main}}\label{sec:3}

As noted in the introduction, Theorem \ref{thm:main} follows from the results of \cite{BoutillierdeTiliere:XORloops} by observing
the following: results of \cite{BoutillierdeTiliere:XORloops} are true for any choice of coupling constants, and
the modified weights $\bar{\Js}$ can be seen as a choice of coupling constants. It is instructive to keep track
of the evolution of the weights through the different steps of the mapping (from the squared Ising model to the bipartite dimer model), 
and to express them as a function of the original coupling constants $\Js$; in particular it is interesting to see the effect of
order and disorder on XOR-Ising configurations.

\subsection{Polygon representation of the squared, modified Ising partition function}\label{sec:31}

A \emph{polygon configuration} of the graph $\Gs$ is a subset of edges $\Ps$ such that every vertex of $\Gs$ is 
incident to an even number of edges of $\Ps$. The set of polygon configurations of $\Gs$ is denoted by $\P(\Gs)$. The set
of polygon configurations $\P(\Gs^*)$ of the dual graph $\Gs^*$ of $\Gs$ is defined similarly.

\begin{figure}[ht]
  \begin{center}
\psfrag{u1}[c][c]{\scriptsize $u_1$}
\psfrag{u2}[c][c]{\scriptsize $u_2$} 
\psfrag{u3}[c][c]{\scriptsize $u_3$} 
\psfrag{u4}[c][c]{\scriptsize $u_4$} 
\psfrag{f1}[c][c]{\scriptsize $f_1$}
\psfrag{f2}[c][c]{\scriptsize $f_2$}
\psfrag{f3}[c][c]{\scriptsize $f_3$}
\psfrag{f4}[c][c]{\scriptsize $f_4$}
\psfrag{f5}[c][c]{\scriptsize $f_5$}
\psfrag{f6}[c][c]{\scriptsize $f_6$}
\psfrag{e}[c][c]{\scriptsize $e$}
\psfrag{ed}[c][c]{\scriptsize $e^*$}
\psfrag{Je1}[c][c]{\scriptsize -$\cosh^{-1}(2\Js_e)$}
\psfrag{Je2}[c][c]{\scriptsize -$\tanh(2\Js_e)$}
\psfrag{Je3}[c][c]{\scriptsize $\tanh(2\Js_e)$}
\psfrag{Je4}[c][c]{\scriptsize $\cosh^{-1}(2\Js_e)$}
\includegraphics[width=\linewidth]{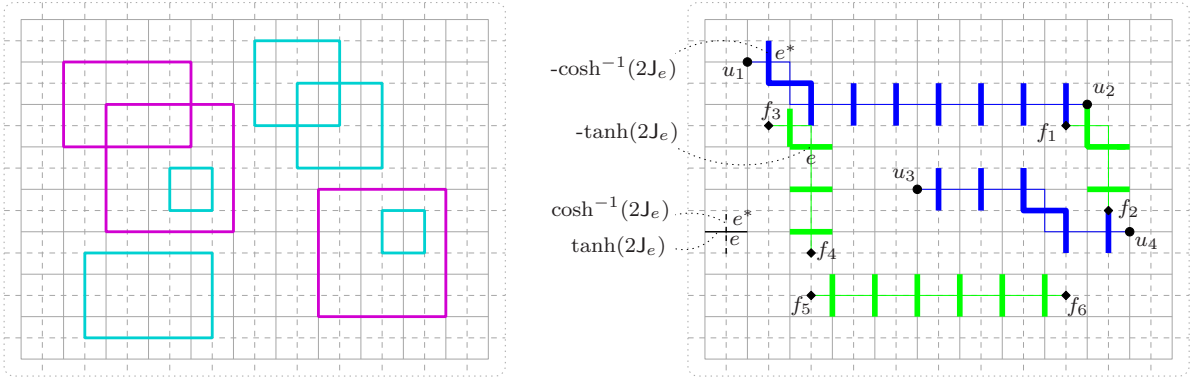}
    \caption{Left: the squared modified Ising partition function can be written as a sum over pairs of non-intersecting
    primal (pink) and dual (turquoise) polygon configurations of $\Gs$ and~$\Gs^*$. Right: modified edge-weights of  
    the graph $\Gs$ and of its dual graph~$\Gs^*$ induced by the modified coupling constants $\bar{\Js}$.}
    \label{fig:FigIsingDefect4}
  \end{center}
\end{figure}

In \cite{BoutillierdeTiliere:XORloops}, we prove that the squared Ising partition function is equal, up to a constant,
to the sum over pairs of non-intersecting polygon configurations of 
the graph $\Gs$ and of its dual graph $\Gs^*$, see Figure \ref{fig:FigIsingDefect4} (left). 
This result is proved for graphs embedded in surfaces of genus $g$, using an idea of Nienhuis \cite{Nienhuis}; it holds 
for any choice of coupling constants. In particular, for graphs embedded in the sphere and for 
the modified weights $\bar{\Js}$, it reads:

\begin{equation}\label{equ:PairPolygon}
[\Zising(\Gs,\bar{\Js})]^2
=\C
\sum_{\{(\Ps,\Ps^*)\in\P(\Gs)\times\P(\Gs^*):\,\Ps\cap \Ps^*=\emptyset\}}
\Bigl(\prod_{e^*\in \Ps^*}\cosh^{-1}(2\bar{\Js}_e)\Bigr)
\Bigl(\prod_{e\in \Ps}\tanh(2\bar{\Js}_e)\Bigr),
\end{equation}
where $\C=2^{|\Vs|+1}\Bigl(\prod_{e\in
\Es}\cosh(2\bar{\Js}_e)\Bigr)$. 

\begin{rem}\label{rem:3}$\,$\vspace{-0.3cm}
\begin{enumerate}
\item From \cite{Nienhuis}, see also \cite{BoutillierdeTiliere:XORloops}, we know that polygon configurations of the dual 
graph $\Gs^*$ arise from the 
low-temperature expansion \cite{KramersWannier1,KramersWannier2} of XOR-Ising configurations:
they are polygon configurations separating clusters of $\pm 1$ spins of the \emph{XOR-Ising model}
\cite{IkhlefRajabpour,Picco,WilsonXOR}, 
obtained by taking the product of the spins
of two independent Ising models with \emph{modified coupling constants}~$\bar{\Js}$.
\item Let us express the modified weights as a function of the original coupling constants $\Js$.
If an edge $e$ of $\Gs$ does not belong to $\Gamma$ or $\Gamma^*$, then
$\tanh(2\bar{\Js}_e)=\tanh(2\Js_e)$; if it belongs to $\Gamma$, then
$\tanh(2\bar{\Js}_e)=\tanh(2\Js_e+i\pi)=\tanh(2\Js_e)$; if it belongs to $\Gamma^*$, then 
$\tanh(2\bar{\Js}_e)=\tanh(-2\Js_e)=-\tanh(2\Js_e)$. If the dual edge $e^*$ of $\Gs^*$ of an edge $e$ of $\Gs$ is such that $e$
does not belong to $\Gamma$ or $\Gamma^*$, then
$\cosh(2\bar{\Js}_e)=\cosh(2\Js_e)$; if $e$ belongs to $\Gamma$, then
$\cosh(2\bar{\Js}_e)=\cosh(2\Js_e+i\pi)=-\cosh(2\Js_e)$; if $e$ belongs to $\Gamma^*$, then 
$\cosh(2\bar{\Js}_e)=\cosh(-2\Js_e)=\cosh(2\Js_e)$. This is summarized in Figure \ref{fig:FigIsingDefect4} (right).
\item The modified coupling constants on $\Gamma^*$ only affect the weight of edges of $\Gs$, and the modified 
coupling constants on $\Gamma$ only affect the weight of edges of the dual graph $\Gs^*$.
\item The term $\prod_{e\in\Es}\cosh(2\bar{\Js}_e)$ in the constant $\C$ of \eqref{equ:PairPolygon} is equal to 
$(-1)^{|\Gamma|}\prod_{e\in\Es}\cosh(2\Js_e)$.
\end{enumerate}
\end{rem}

Since the proof \cite{BoutillierdeTiliere:XORloops} of Equation \eqref{equ:PairPolygon}
greatly simplifies in the genus 0 case, we give it here to make this
paper self contained.

\begin{proof}
The squared modified partition function is equal to:
\begin{equation*}
[\Zising(\Gs,\bar{\Js})]^2=\sum_{\sigma,\sigma'\in\{-1,1\}^\Vs}\prod_ {
e=uv\in
\Es}e^{\bar{\Js}_e\sigma_u\sigma_v}e^{\bar{\Js}_e\sigma_u'\sigma_v'}. 
\end{equation*}

For every pair of spin configurations $\sigma,\sigma'$, denote by $\tau$ the XOR-Ising configuration, obtained 
by taking the product $\sigma\sigma'$: $\forall u\in \Vs,\;\tau_u=\sigma_u\sigma_u'$, then $\tau\in\{-1,1\}^\Vs$. 

Since $\sigma$ and $\sigma'$ take values in $\{-1,1\}$, we have
$\sigma'=\tau\sigma$, and the squared modified partition
function can be written as:
\begin{equation*}
[\Zising(\Gs,\bar{\Js})]^2
=\sum_{\tau,\sigma\in\{-1,1\}^\Vs}
\Bigl(\prod_{e=uv\in \Es}
e^{\bar{\Js}_e\sigma_u\sigma_v(1+\tau_u\tau_v)}\Bigr).
\end{equation*}
Let us now fix a XOR-spin configuration $\tau$. Denote by 
$\Vs_\tau^1,\cdots,\Vs_\tau^{k_\tau}$ the partition of vertices of $\Vs$
corresponding
to clusters of $\pm 1$ spins of $\tau$. For every $\l\in\{1,\cdots,k_{\tau}\}$,
let
$\Es_\tau^\l$ be the subset of edges joining vertices of $\Vs_\tau^\l$. Denote by
$\Es_\tau=\cup_{\l=1}^{k_\tau}\Es_\tau^\l$, and by $(\Es_\tau)^c=\Es\setminus
\Es_\tau$. Then, 
for every $e=uv\in \Es_\tau$, $\tau_u\tau_v=1$, and for every $e\in (\Es_\tau)^c$,
$\tau_u\tau_v=-1$, implying that:
\begin{equation*}
[\Zising(\Gs,\bar{\Js})]^2
=\sum_{\tau\in\{-1,1\}^\Vs}
\sum_{\sigma\in\{-1,1\}^\Vs}
\prod_{e=uv\in \Es_\tau} e^{2\bar{\Js}_e\sigma_u\sigma_v}.
\end{equation*}
Exchanging the sum over spins $\sigma$'s and the product over edges of $\Es_\tau$ yields:
\begin{equation}\label{equ:mixedcontour0}
[\Zising(\Gs,\bar{\Js})]^2
=\sum_{\tau\in\{-1,1\}^\Vs}
\prod_{\l=1}^{k_\tau}
\Bigl[
\sum_{\sigma^\l\in\{-1,1\}^{\Vs^{\l}_\tau}}
\Bigl(\prod_{e=uv\in \Es^{\l}_\tau}
e^{2\bar{\Js}_e\sigma_u\sigma_v}\Bigr)\Bigr].
\end{equation}
That is, for every $\l\in\{1,\cdots,k_\tau\}$, we have the partition function of
an Ising model on $\Gs^{\l}_\tau=(\Vs^{\l}_\tau,\Es^{\l}_\tau)$, with modified, doubled
coupling constants $2\bar{\Js}$. Using Kramers and Wannier high temperature expansion \cite{KramersWannier1,
KramersWannier2} for each of these modified Ising models, we obtain:
\begin{align}\label{equ:mixedcontour4}
\prod_{\l=1}^{k_\tau}
\Bigl[
\sum_{\sigma^\l\in\{-1,1\}^{\Vs^{\l}_\tau}}
\Bigl(\prod_{e=uv\in \Es^{\l}_\tau}
e^{2\bar{\Js}_e\sigma_u\sigma_v}\Bigr)\Bigr]&=
\prod_{\l=1}^{k_\tau}\Bigl[2^{|\Vs_\tau^\l|}
\Bigl(\prod_{e\in\Es_\tau^\l}\cosh(2\bar{\Js}_e)\Bigr)
\sum_{\Ps\in\P(\Gs^\l_\tau)}
\Bigl(\prod_{e\in \Ps}\tanh(2\bar{\Js}_e)\Bigr)
\Bigr]=\nonumber\\
&=2^{|\Vs|}
\Bigl(\prod_{e\in\Es_\tau}\cosh(2\bar{\Js}_e)\Bigr)
\prod_{\l=1}^{k_\tau}
\Bigl[
\sum_{\Ps\in\P(\Gs^\l_\tau)}\Bigl(\prod_{e\in \Ps}\tanh(2\bar{\Js}_e)\Bigr)\Bigr].
\end{align}
Plugging \eqref{equ:mixedcontour4} into the squared modified partition function 
\eqref{equ:mixedcontour0} yields,
\begin{equation*}
[\Zising(\Gs,\bar{\Js})]^2
=\C' \sum_{\tau\in\{-1,1\}^{\Vs}}
\bigl(\prod_{e\in(\Es_\tau)^c}\cosh^{-1}(2\bar{\Js}_e)\bigr)
\prod_{\l=1}^{k_\tau}
\Bigl(\sum_{\Ps\in\P(\Gs^\l_\tau)}\prod_{e\in \Ps}\tanh(2\bar{\Js}_e)\Bigr),
\end{equation*}
where $\C'=2^{|\Vs|}\Bigl(\prod_{e\in\Es}\cosh(2\bar{\Js}_e)\Bigr)$.

The proof is concluded by assigning, as in the low temperature expansion, dual polygon configurations
separating clusters of spins of XOR-Ising
configurations. Note that the constant $\C'$ and the constant $\C$ of the statement differ by a factor 2
because two spin configurations are assigned to a given dual polygon configuration.
\end{proof}

\subsection{Bipartite dimer representation of the polygon representation}\label{sec:32}

\begin{figure}[ht]
  \begin{center}
\psfrag{e}[c][c]{\scriptsize $e$}
\psfrag{ed}[c][c]{\scriptsize $e^*$}
\psfrag{w1}[c][c]{\scriptsize $\tanh(2\bar{\Js}_e)$}
\psfrag{w4}[c][c]{\scriptsize $\cosh^{-1}(2\bar{\Js}_e)$}  
\includegraphics[width=\linewidth]{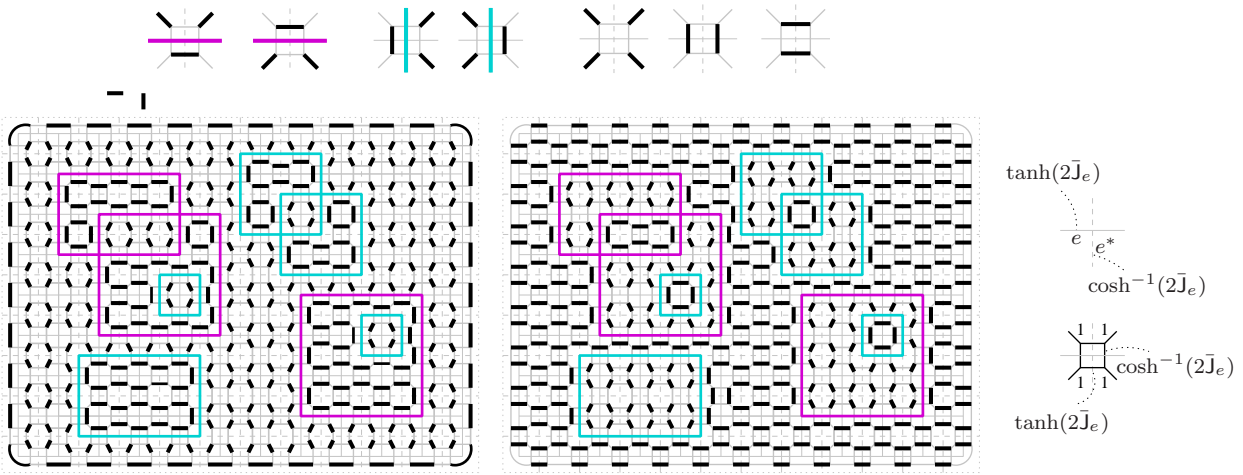}
    \caption{Mapping between pairs of non-intersecting primal and dual polygon configurations
    of $\Gs$ and $\Gs^*$ and dimer configurations of $\GQ$ \cite{Nienhuis,WuLin,Dubedat}. Top: mapping on the local level. Bottom: mapping on the global level. 
    Given a pair of non-intersecting primal and dual polygon configurations, there are two possible leg configurations for the
    corresponding dimer configuration; then the configuration of quadrangles with 2 or 4 matched legs is fixed, and quadrangles having 0
    matched leg each has two possible dimer configurations. Right: mapping of the weights.}\label{fig:FigIsingDefect5}
  \end{center}
\end{figure}

We proceed as in \cite{BoutillierdeTiliere:XORloops}. The sum over pairs of non-intersecting primal and dual polygon configurations
naturally maps to a 6-vertex model \cite{Nienhuis}. This 6-vertex model is free-fermionic when polygon edge-weights arise from
two independent Ising models (because $[\cosh^{-1}(2\bar{\Js}_e)]^2+[\tanh(2\bar{\Js}_e)]^2=1$). 
The free-fermionic 6-vertex model then maps to the dimer model on the graph $\GQ$ defined in 
Section~\ref{sec:23} \cite{WuLin, Dubedat}. The mapping from pairs of non-intersecting primal and dual polygon configurations of $\Gs$ and
$\Gs^*$, to dimer
configurations of $\GQ$ can be explained without going through the 6-vertex model. It is summarized in 
Figure~\ref{fig:FigIsingDefect5}.

As a consequence, we obtain:
\begin{equation}\label{equ:BipartiteDimer}
\sum_{\{(\Ps,\Ps^*)\in\P(\Gs)\times\P(\Gs^*):\,\Ps\cap \Ps^*=\emptyset\}}
\Bigl(\prod_{e^*\in \Ps^*}\cosh^{-1}(2\bar{\Js}_e)\Bigr)
\Bigl(\prod_{e\in \Ps}\tanh(2\bar{\Js}_e)\Bigr)=\frac{1}{2}\Zdimer(\GQ,\nu(\bar{\Js})),
\end{equation}
where the dimer weight function $\nu(\bar{\Js})=(\nu(\bar{\Js})_{\es})_{\es\in\EQ}$ is given by: 
\begin{equation*}\label{equ:dimerweights}
\nu(\bar{\Js})_{\es}= 
\begin{cases}
1&\text{ if $\es$ is a leg}\\
\tanh(2\bar{\Js}_{e})&\text{ if $\es$ is ``parallel'' to a primal edge $e$ of $\Gs$}\\
\cosh^{-1}(2\bar{\Js}_{e})&\text{ if $\es$ is ``parallel'' to the dual edge $e^*$ of an edge $e$ of $\Gs$}.
\end{cases}
\end{equation*}

Using the computations of the weights of Section \ref{sec:31}, we can express the dimer weight function $\nu(\bar{\Js})$ as a function
of the original coupling constants, see also Figure \ref{fig:FigIsingDefect2} (right):
\begin{equation}\label{equ:modifieddimerweights}
\nu(\bar{\Js})_{\es}= 
\begin{cases}
1&\text{ if $\es$ is an external edge}\\
\tanh(2\Js_{e})&\text{ if $\es$ is ``parallel'' to an edge $e$ of $\Gs$, $e\notin\Gamma^*$}\\
-\tanh(2\Js_{e})&\text{ if $\es$ is ``parallel'' to an edge $e$ of $\Gs$, $e\in\Gamma^*$}\\
\cosh^{-1}(2\Js_{e})&\text{ if $\es$ is ``parallel'' to the dual edge $e^*$ of an edge $e$ of $\Gs$, $e\notin\Gamma$}\\
-\cosh^{-1}(2\Js_{e})&\text{ if $\es$ is ``parallel'' to the dual edge $e^*$ of an edge $e$ of $\Gs$, $e\in\Gamma$}.
\end{cases}
\end{equation}

Combining Equations \eqref{equ:PairPolygon}, \eqref{equ:BipartiteDimer}, Point 3 of Remark \ref{rem:3}, and
using the fact that the sequence of mappings works for all choices of coupling constants, we obtain:
\begin{align*}
[\Zising(\Gs,\Js)]^2&=2^{|\Vs|}\bigl(\prod_{e\in
\Es}\cosh(2\Js_e)\bigr)\,\cdot\,\Zdimer(\GQ,\nu(\Js))\\
[\Zising(\Gs,\bar{\Js})]^2&=2^{|\Vs|}(-1)^{|\Gamma|}\bigl(\prod_{e\in
\Es}\cosh(2\Js_e)\bigr)\,\cdot\,\Zdimer(\GQ,\nu(\bar{\Js})).
\end{align*}

Taking the ratio yields Theorem \ref{thm:main}.

\section{Consequences}\label{sec:4}

As a consequence of Theorem \ref{thm:main}, we obtain expressions as ratio of bipartite dimer partition functions for squared
quantities of interest in the study of the Ising model. Throughout this section, we use the notations of Sections \ref{sec:1}, 
\ref{sec:2} and \ref{sec:3}.

First, following \cite{KadanoffCeva} we consider $2n$-\emph{spinor variables}. This amounts
to taking $m=n$ and choosing $u_j$, $f_j$ in such a way that
$u_j$ is on the boundary of the face of $\Gs$ defined by the dual vertex $f_j$. Then, specifying Theorem \ref{thm:main} to this choice of
vertices yields an expression for squared spinor variables correlations as the ratio of bipartite dimer partition functions.

Next, let us consider \emph{$2n$-spin correlations} $\EE[\sigma_{u_1}\dots\sigma_{u_{2n}}]$. This enters the framework of this paper by taking
$\Gamma^*$ to be the empty set. More precisely, by Kadanoff and Ceva \cite{KadanoffCeva}, $2n$-spin correlations are equal to:
\begin{equation*}
\EE[\sigma_{u_1}\dots\sigma_{u_{2n}}]=(-i)^{|\Gamma|}\frac{\Zising(\Gs,\bar{\Js})}{\Zising(\Gs,\Js)}=
(-i)^{|\Gamma|}\mathord{<}\sigma_{u_1}\dots\sigma_{u_{2n}}\mathord{>}_{(\Gs,\Js)},
\end{equation*}
where 
\begin{equation*}\label{equ:modifiedIsingweightsspins}
\bar{\Js}_e=
\begin{cases}
\Js_e+i\frac{\pi}{2}&\text{ if }e\in\Gamma\\
\Js_e&\text{ otherwise}.
\end{cases}
\end{equation*}
Computing the corresponding dimer weight function $\nu(\bar{\Js})$ using Equation \eqref{equ:modifieddimerweights} yields:
\begin{equation}\label{equ:modifieddimerweightsspins}
\nu(\bar{\Js})_{\es}= 
\begin{cases}
1&\text{ if $\es$ is an external edge}\\
\tanh(2\Js_{e})&\text{ if $\es$ is ``parallel'' to an edge $e$ of $\Gs$}\\
\cosh^{-1}(2\Js_{e})&\text{ if $\es$ is ``parallel'' to the dual edge $e^*$ of an edge $e$ of $\Gs$, $e\notin\Gamma$}\\
-\cosh^{-1}(2\Js_{e})&\text{ if $\es$ is ``parallel'' to the dual edge $e^*$ of an edge $e$ of $\Gs$, $e\in\Gamma$}.
\end{cases}
\end{equation}
As a consequence of Theorem \ref{thm:main}, we obtain the following.
\begin{cor}\label{cor:main}
The squared $2n$-spin correlations $\EE[\sigma_{u_1}\dots\sigma_{u_{2n}}]^2$ is the following ratio of bipartite dimer partition
functions:
\begin{equation*}
\EE[\sigma_{u_1}\dots\sigma_{u_{2n}}]^2=\frac{\Zdimer(\GQ,\nu(\bar{\Js}))}{\Zdimer(\GQ,\nu(\Js))},
\end{equation*}
where the dimer weight function $\nu(\bar{\Js})$ is given by Equation \eqref{equ:modifieddimerweightsspins}.
\end{cor}

Finally, let us express the \emph{magnetization} $\EE^+[\sigma_u]$ which is the expectation of a single spin $u$ under plus-boundary conditions 
(if free boundary conditions were considered, the magnetization would be equal to zero by symmetry). Recall that fixing plus-boundary amounts to
taking all spins on boundary vertices of a face $\Fs$ of $\Gs$ to be $+1$. Magnetization enters the framework of this paper
by taking $\Gamma^*$ to be the empty set, $\Gamma$ to be a single path $\gamma$, and by using the procedure of Section \ref{sec:22} for 
treating plus-boundary conditions. More precisely, by \cite{KadanoffCeva}, the magnetization is equal to:
\begin{equation*}
\EE^+[\sigma_u]=(-i)^{|\gamma|} \frac{\Zising^+(\Gs,\bar{\Js})}{\Zising^+(\Gs,\Js)},
\end{equation*}
where $\Zising^+(\Gs,\bar{\Js})$ is the modified, plus-boundary condition Ising partition function, modified along a single path $\gamma$
of the graph $\Gs$, where $\gamma$ joins a vertex on the boundary of $\Fs$ to the vertex $u$; this quantity is independent of the choice of boundary
vertex. Let us suppose that $\gamma$ does not use edges on the boundary of $\Fs$, and let $\Gs'$ be the graph obtained from $\Gs$ by merging
the face $\Fs$ into a single vertex $v$, see Figure \ref{fig:FigBoundary2} (left). Note that in $\Gs'$, the path $\gamma$ joins the vertices $v$ and $u$.
Using the argument of Section \ref{sec:22} for handling plus-boundary conditions, we obtain:
\begin{equation*}
\frac{\Zising^+(\Gs,\bar{\Js})}{\Zising^+(\Gs,\Js)}=\frac{\frac{1}{2}\bigl(\prod_{e\in\Es_{\partial \Fs}}e^{\Js_e}\bigr)
\Zising(\Gs',\bar{\Js})}{\frac{1}{2}\bigl(\prod_{e\in\Es_{\partial \Fs}}e^{\Js_e}\bigr)\Zising(\Gs',\Js)}=
\frac{\Zising(\Gs',\bar{\Js})}{\Zising(\Gs',\Js)}.
\end{equation*}
Since $\gamma$ does not use boundary edges of the face $\Fs$, it has the same number of edges in the graphs $\Gs$ and $\Gs'$. 
We have thus proved the following.
\begin{lem}
The magnetization $\EE^+[\sigma_u]$ in the graph $\Gs$, is equal to the pair-spin correlations $\EE[\sigma_u\sigma_v]$ in the
graph $\Gs'$.
\end{lem}
As a consequence, the expression as ratio of bipartite dimer partition functions for the squared magnetization is a specific case of Corollary 
\ref{cor:main}.


\bibliographystyle{alpha}
\bibliography{survey}

\end{document}